# Title:
# Reverse enGENEering of regulatory networks from Big Data: a guide for a biologist

Xiaoxi Dong[1], Anatoly Yambartsev[4], Stephen Ramsey[2], Lina Thomas[4], Natalia Shulzhenko[3], Andrey Morgun[1]

Affiliations:

1. College of Pharmacy, Oregon State University, Corvallis, Oregon, United States

2. School of Electrical Engineering and Computer Science, Biomedical department, Oregon State University, Corvallis, Oregon, United States

3. College of Veterinary Medicine, Biomedical department, Oregon State University, Corvallis, Oregon, United States

4. Department of Statistics, Institute of Mathematics and Statistics, University of Sao Paulo, Sao Paulo, SP, Brazil.

Xiaoxi Dong: Postdoctoral scholar

Andrey Morgun: Assistant professor

Stephen Ramsey: Assistant professor

Natalia Shulzhenko: Assistant professor

Anatoly Yambartsev: Associate professor

Lina Thomas: Graduate student

Correspondence to: andriy.morgun@oregonstate.edu; anemorgun@hotmail.com

## Abstract
Omics technologies enable unbiased investigation of biological systems through massively parallel sequence acquisition or molecular measurements, bringing the life sciences into the era of Big Data. A central challenge posed by such omics datasets is how to transform this data into biological knowledge. For example, how to use this data to answer questions such as: which functional pathways are involved in cell differentiation? Which genes should we target to stop cancer? Network analysis is a powerful and general approach to solve this problem consisting of two fundamental stages, network reconstruction and network interrogation. Herein, we provide an overview of network analysis including a step by step guide on how to perform and use this approach to investigate a biological question. In this guide, we also include the software packages that we and others employ for each of the steps of a network analysis workflow.

# Introduction:

In saying that we understand a biological process we usually mean that we are able to predict future events and manipulate the process guiding it into a desirable direction. Thus, biological inquiry could be viewed as an attempt to understand how a biological system transits from one state to another. Such transitions underlie a wide range of biological phenomena from cell differentiation to recovery from disease.

In attempting to understand these transitions, a simple and frequently used approach is to compare two states of a system (e.g. before and after stimulus, with and without mutation, or healthy and diseased).

Although more sophisticated approaches with time series data, dose-effect data, or three or more classes data can be also used, this paper is devoted to the two-classes/states study design. Furthermore, most of the methods that we describe can be used for other study designs with slight modifications.

Today, omics technologies enable unbiased investigation of biological systems through massively parallel sequence acquisition or molecular measurements, bringing the life sciences into the era of Big Data. A central challenge posed by such omics datasets is how to navigate through the "haystack" of measurements (for example, differential expression between two states) to identify the "needles" comprised of the critical causal factors.

A powerful and general approach to this problem is network analysis and its two fundamental stages, network reconstruction and network interrogation. For omics molecular measurements such as gene expression, covariation networks have become a dominant paradigm in the field of network analysis. Multiple groups including ours have been successfully using such methods to gain a systems-level understanding of biological processes and to reveal mechanisms of different diseases.[1] However, due to the rapid pace of evolution of techniques and omics datasets, the practical application of network analysis has traditionally required a dedicated computational biologist. This requirement has limited the extent to which the larger biological sciences community has benefited from network analysis.

Herein, we provide an overview of covariation network reconstruction and interrogation including a step by step guide on how to perform and use network analysis to investigate a biological question (Figure 1).

In this guide, we include the software packages that we employ (and specific pointers to the methods/software used by other groups) for each of the steps of a model network analysis workflow. Although in this guide we mostly focus on covariation networks, the analysis steps related to network interrogation are applicable to other types of networks such as semantic networks or molecular interaction networks.

**The guide:**

In general, the types of omics measurements that are amenable to network analysis range from microarrays to next-generation sequencing (for genotyping or transcriptome profiling) and mass spectrometry-based proteomics and metabolomics data. While network analysis is usually and most straightforwardly applied to one type of omics data at a time (i.e. a homogeneous dataset), integrative networks are becoming more popular under the premise that the resulting networks more comprehensively reflect the underlying biology. In this guide we focus on gene expression data to illustrate the process of network reconstruction and interrogation.

**Network reconstruction**

The first stage of network analysis is network reconstruction, which is the data-driven discovery or inference of the entities/nodes (transcripts, proteins, genes, metabolites, or microbes) and relationships/edges between these entities that together constitute the biological network. Each type of 'omics' measurement technology has a specific procedure for reducing the raw data to a consensus abundance or frequency measure for each entity. Here, we describe the steps involved in network reconstruction starting from entity abundance/frequency data.

*Normalization*

Customarily, abundance data are normalized in order to correct for sample-to-sample variation in the overall distribution of abundance values (or more generally, to normalize specific quantities that depend on the distribution). Measurements of gene expression levels (as well as other types of omics data) can be affected by a variety of non-biological factors including unequal amount of starting RNA, different extent of labeling or different efficiencies of detection between samples. Before normalization, data is often log-transformed in order to stabilize variances when measurements span orders of magnitude.

Frequently-used normalization schemes include median normalization, quantile normalization, LOWESS normalization[2] for RNA microarray data, RPKM,[3] trimmed mean of M-values,[4] and DESeq[5] for RNA-seq data. In practice, we use normalization procedures available in the software package BRB ArrayTools(Richard Simon and the BRB-Arraytools development team, unpublished, 2004) for normalization of microarray data (Table1). In addition, most normalization procedures are available as packages in the Bioconductor toolkit.[6] Systematic evaluations of transcriptome normalization methods have been reported for both microarrays[5] and RNA-seq,[7] however, evaluations using large numbers of sample groups are needed in order to determine which normalization method is most appropriate for covariance network inference. Selection of an appropriate normalization method is clearly important, given that selection of a suboptimal normalization scheme can lead to over-estimation of gene-gene correlation coefficients.[5]

Beyond transcriptome profiling, different omics data types may benefit from different types of normalization. For example, new methods have been proposed for metabolomics[8] and microbiome[9] data normalization.

Although there is no consensus about a best method for all types of data, in the experience of the authors,[10-15] simple methods such as quantile, lowess or even median normalization perform reasonably well for class comparison and correlation if there are no major biases in the data such as batch effects.

*Discovery of differentially expressed genes*

A crucial step in network reconstruction is the identification of the relevant subset of variables that will constitute the "nodes" in the network; for a transcriptome profiling study, these would be genes for which there is significant differential expression between the sample groups. A variety of statistical tests are commonly used for the identification of differentially expressed genes (DEGs), including Welch's *t*-test, moderated *t*-test, and permutation tests. For parametric tests, accurate estimation of intra-sample-group variance is a critical issue; two improved variance estimation techniques are the locally pooled error and empirical Bayes methods. Because omics data analysis typically involves tens of thousands of statistical tests, the correction for multiple hypothesis is essential.[16] To find DEGs, we usually use the t-test with the ordered set of *P* values converted to cumulative false discovery rate (FDR) estimates, for which a typical cutoff would be 10%. Both statistical functions are implemented in BRB-ArrayTools.[17]

During the last two decades multiple statistical approaches have been proposed for differential expression testing.[18] Overall they provide similar results with small differences.[18] Thus careful study design (rather than "trash in, trash out") and the use of meta-analysis techniques to integrate multiple datasets, can be more important for reliable DEGs discovery than a choice of one or another statistical test.

*Finding links between nodes (correlation analysis for network reconstruction)*

The central principles underlying correlation network analysis are (1) that DEGs reflect functional changes and (2) that DEGs do not work individually, rather the interactions between multiple DEGs result in functional alteration of the biological system. In gene expression networks, nodes represent genes and edges represent significant pairwise associations between gene expression profiles. To reconstruct the network, the Pearson or Spearman correlation coefficient can be used as an association (similarity) measure for each possible pair of DEGs, with a cutoff for statistical significance (an FDR cutoff of 10% for the $n(n-1)/2$ possible pairwise associations tested) and for a minimum correlation level. Together with the nodes, the edges that pass this procedure constitute a network. Note that correlations should be calculated within a group of samples that belong to one class/biological state (pooling samples from different states/classes to compute the correlation coefficient leads to significant bias).

In practice, normalized expression data for DEGs are retrieved and pairwise correlations are calculated for each class (biological state) separately using the R statistical analysis software, with the function `cor.test`; FDR is calculated using the function `p.adjust`.

Several other software programs that can be used for calculating gene-gene associations (correlations, mutual information and others) are represented in Table 1.

*Discriminating between direct and indirect links*

Co-variation networks in general consist of connections that result from a combination of direct and indirect effects between variables/nodes. For example, if a variable Y strongly depends on variable X and variable Z also depends on X, it is likely that a high 5association (e.g. correlation) will exist between Y and Z even if there is no direct dependence between them (Figure 2). Moreover, even if a true dependence exists between a pair of variables/nodes, its strength estimation can be biased by additional indirect relationships.[19] For this reason, correlation networks in general have many edges that reflect indirect relationships between pairs of genes, where no direct relationship exists.

Direct effects can be defined as the association between two variables holding the remaining variables constant.[20] Usually all effects that are not direct are called an *indirect effects.* The identification of direct links in a network is a one of important goals in reverse-engineering.

To infer direct links between DEGs, we have been using the partial correlation coefficient.[21,22] To calculate partial correlations we use a method called the Inverse Method.[23] Its implementation is straightforward in R using the function `cor2pcor` from package "`corpcor`". The detailed algorithm is described in supplementary text 1. After calculation of partial correlation, the network can be built using links with absolute value of partial correlation larger than user defined threshold.

Several other methods have been proposed to discriminate between direct and indirect links in covariation networks.[24-27] We have also suggested using a variant of the partial correlation that we call the Local Partial Correlation in order to overcome the limitations of other methods.[28]

*PUC (Proportion of Unexpected Correlations)*
Recently, our group has proposed a new method that allows identifying and removing approximately half of erroneous edges from a covariation.[29] The method, called Proportion of Unexpected Correlations (PUC), also provides an estimate of the proportion of erroneous edges. The method takes into account a relation between direction of regulation of two DEGs and the sign of correlation between the two genes. Thus, two up- and two down-regulated genes must correlate positively; and a pair of oppositely regulated genes (one upregulated and one downregulated) should have negative correlation. Any deviation from this rule represents unexpected/erroneous edges and is removed from the network (Figure 3). The proportion of these unexpected edges provides an error estimates for a whole network. For network reconstruction, each edge in a network can be evaluated and removed if it is unexpected.

*Meta-analysis*
In omics-based network reconstruction, because of the large number of variables (up to tens of thousands) and the limited number of samples (tens or hundreds in the best case scenario), it is critical to assess the reproducibility of results. Although widely used methods (e.g., FDR[30]) enable accounting for multiple hypothesis tests, the discrepancy between the number of samples and variables inherent to omics datasets limits the sensitivity and specificity for detecting edges through network reconstruction. One solution to this problem is to employ meta-analysis, a statistical approach for combining results of different studies[31] (in order to achieve high reproducibility) and leverage omics datasets from many studies, obtained from standardized omics data repositories.

Good examples of such repositories are Gene Expression Omnibus (GEO)[32] and Array Express[33] (for transcriptomics and epigenomics datasets); PRIDE[34] (for proteomics datasets), the Human Metabolome Database[35] (for metabolomics datasets), and lipid MAPS[36] (for lipidomics datasets). Additionally, molecular interaction data from the BioGRID[37] or BioCyc databases[38] can be used as a prior for edge reconstruction.
In meta-analysis of multiple datasets – whether from publicly available datasets or experiments produced in the same lab – the strategy is usually the same. Datasets should be selected based on their congruence with the central biological question of the meta-analysis, and they should pass some predefined sample size and quality requirements (e.g., number of measured/detected genes).
Multiple approaches have proposed for meta-analysis of gene expression data (Table 1).[39,40] In supplementary text 1 we describe a simple algorithm that we have employed for integrating differential expression, correlations, and differential associations/correlations.[11]

*Differentially co-expressed gene pairs*

The networks discussed above model pan-dataset ("static") correlations between genes that change their expression when the biological system transits from one state to another. However, the sets of edges within a gene co-variation network can themselves vary from state to state, for example, when two genes are highly correlated in a subset of conditions but not across all conditions.[41] Such a gene pair is called a differentially co-expressed gene pair (Figure 4).
It has been shown that differentially co-expressed gene pairs frequently play critical roles in pathogenesis. Several studies have explored gene coexpression changes in cancer, revealing known cancer genes that were on top-ranked among co-expression changes, but not necessary (separately) among differentially expressed genes.[15,42]
In order to search for differentially co-expressed gene pairs our group adapted a simple approach called Differentially Associated Pairs (DAPs).[15] The DAPs algorithm is described in supplementary text 1.
In addition to DAPs, multiple methods/software have been developed to find changing edges in gene expression networks.[15,43] (Table 1)

*Integrating heterogeneous omics data types: <u>inter-omics</u> networks*

The integration of different omics data types holds great promise for enabling more robust network reconstruction and detection of causal interactions in a particular biological context. For example, genome-wide measurements of epigenetic marks and transcriptome data can be combined to elucidate mechanisms of gene regulation.[44-46] In cancer bioinformatics, integration of gene copy number data (chromosomal aberrations) and gene expression measurements has enabled the discovery of key drivers.[11,47] And integration of metagenomics data from gut microbiota with intestinal gene expression reveals new mechanisms of crosstalk between microbes and their hosts (Andrey Morgun et.al. 2014 unpublished data).

Approaches for omics data integration generally fall into one of two modalities: first (and most prevalent) is integrating different types of data generated for a given gene/gene product.[48] In other words, a given node pertains to more than one network (e.g. measurements of the copy numbers of gene A and transcript levels of gene A pertain to genomic and transcriptomic networks respectively) (Figure 5A) .

The other type of integration makes an edge/link between two nodes from different omics networks. We call the result of such integration an inter-omics network. An inter-omics network is a bipartite network in which each edge connects two nodes of different omics types (Figure 5B). There are two different approaches how to infer such inter-omics links/edges. First one is based on bringing into reconstruction an experimental result supporting a link between nodes of different omics. For example, nodes from proteomics and metabolomics networks can be connected based on the experiment showing that a specific protein is an enzyme necessary for the production of a given metabolite. Second approach that infers edges between different omics establishes connection between two different (knowledge-wise unrelated) quantitative variables based on their statistical association (e.g. correlation between gene expression and abundance of metabolites). Thus, the entire reconstruction procedure consists of inference on networks of each omics type separately and then integration of these two networks into the inter-omics network. This is a straightforward and easily implementable algorithm. Furthermore, there is a popular tool, integrOmics, that is used for heterogeneous data integration using partial least squares regression.[49]

**Network interrogation**

To gain maximal insights from a biological network that has been reconstructed as described above, systematic analysis of the network ("network interrogation") is essential. In this section we describe several network interrogation techniques for investigating specific types of biological questions.

*Revealing potential mechanisms of a biological process or disease*

This goal is achieved by identification of pathways involved in the process, key regulatory nodes of those pathways, and interactions between identified pathways (including identification of nodes in network responsible for the interaction).

*Which functional pathways are involved?*

Finding dense subnetworks (i.e. modules)

From a functional standpoint, subsets of genes that are highly interconnected in the correlation network (modules[50]) are often involved in similar biological processes. Tools for identification of modules include MCODE,[51] cfinder,[52] and graph clustering (MCL).[53]

A key advantage of network module analysis (vs. direct clustering of genes from the data) is that while modules would include genes up- and down-regulated that correspond to potential stimulatory and inhibitory relations within a given functional pathway, traditional clustering approaches would group genes with similar behavior, thus separating up- and down-regulated genes from the same pathway into different clusters.

Enrichment analysis with external data (e.g. Gene Ontology)

Once genes that work together (modules) are identified the next step is to infer their biological function. This is usually performed by using literature-curated, gene-centric biological knowledge bases that connect genes to Gene Ontology (and other types of) functional categories ("terms"). If a module is enriched for genes that are associated with a particular biochemical pathway, a location in a genome, or a location in cellular compartment, that finding can provide a basis for a hypothesis about the function of the module.
A plethora of tools are available for gene functional enrichment analysis (Table 1). For example, gene sets can be annotated by pathways using tools like SubpathwayMiner[54] or by gene ontology terms using tools like DAVID.[55] Various tools can further construct a functional network based on gene-gene functional associations: Bingo[56] and EnrichmentMap.[57]

*Key regulators of pathways/modules*

Identifying key molecular regulators of the biological response or system under study is often a primary goal in omics studies, especially those with a tractable cellular model where molecular or genetic perturbations can be introduced. There are two major complementary strategies for finding key regulators in covariation networks: (i) using network topological properties and (ii) incorporating additional data into networks that provides information about causes of regulation for some nodes in a network.
Topological properties that have been described to date as pointing to key regulators mostly define different flavors of connectivity of a node. Those properties are: degree and centrality measures such as betweenness centrality, closeness centrality, eigenvector centrality, etc. In recent work, nodes with high betweenness centrality (so-called "bottlenecks") were shown to be predictive of gene essentiality.[58] For example, such topological characteristics have been found to be associated with genes that are critical for pathogen virulence,[59] and with genes that are targets for hepatitis C virus.[60] The estimation of these parameters is a straightforward and be easily accomplished using the Cytoscape plug-in called "NetworkAnalyzer".[61] Importantly, these properties need not be analyzed in isolation but can complement another approach we discuss below.[62]

Integrating additional information in order to find causes of regulation
It is axiomatic that a gene-gene network that has been reconstructed based on correlation analysis does not discriminate between direct regulation and "common cause".[63] Therefore, it is common to incorporate into covariation network several types of complementary biological data that can directly or indirectly indicate that one gene regulates another.[64,65] Overlaying such information on a co-expression network one can establish the directionality of some edges that improves precision of identification of key regulators. The types of biological information include genetic variants (aberrations, mutations, gene polymorphisms etc.), epigenetic modifications, transcription factors and other types of gene expression regulation such as miRNA. For example, integrating genomic aberrations with global gene expression led to the discovery of key drivers of melanoma,[47] and breast[66] and cervical cancers.[11] Similarly, eQTLs (Expression Quantitative Trait Loci) were integrated with networks associated with diabetes and obesity, revealing causal genes of specific molecular pathways operating in these diseases.[67]

Integration of information about binding sites (or computationally predicted binding sites) of transcription factors into covariation networks is a particularly powerful approach,[68] because the direction of causality for a connection between a transcription factor and a target gene is presumed to be known. While computational analysis of transcription factor binding site (TFBS) databases (such as TRANSFAC) can suggest the possibility of regulation by a given transcription factor, omics approaches for identification transcription factor binding sites such as ChIP-Seq provide more definitive genome-wide location information for the transcription factor in an investigated sample. The directionality information provided by those methods can be incorporated into network interrogation to generate more accurate prediction of key regulators.[69] MicroRNAs (miRNAs) are another important class of gene expression regulators that modulate (primarily down-regulate) expression of target genes either by inhibition of translation or promoting mRNA degradation. In the past few years, ~1,881 miRNA genes have been identified in humans (according to miRBase, http://www.mirbase.org/cgi-bin/mirna_summary.pl?org=hsa), and knowledge of miRNA-target interactions is accumulating both by experimental validation and computational prediction.[70,71] More accurate genome-wide miRNA target sequence location information allows the possibility of generating a miRNA-mRNA regulatory network, which could provide a more complete view of regulatory relationship in biological process. In a recent work, Sumazin et.al. integrated gene and miRNA expression data from sample matched datasets and constructed a comprehensive miRNA-gene interaction network, inferring that PTEN is a key regulator of gliomagenesis.[62]

Integrating multiple types of data simultaneously can increase the precision of computational predictions. For example, one of us has reported that "using motif scanning and Histone acetylation local minima, improves the sensitivity for TF binding site prediction by approximately 50% over a model based on motif scanning alone".[44] In another work, Yang et.al integrated gene expression with gene copy number alternation, DNA methylation, associated miRNA expression and miRNA target prediction to identify key regulatory miRNA genes that regulate ovarian cancer development and then experimentally validated the function of one predicted miRNA gene.[72] In practice, multiple tools have been developed for the integration of different resources of information to infer network and/or identification of key regulators (Table 1).

*How the pathways interact*

As networks represent models of global changes in biological system they usually contain several groups of genes exerting specific biological functions. Cooperation of these functions/pathways plays an important role in regulating biological processes.[68] Transcriptional network can be viewed as a group of interacting pathways/modules (meta-modules) rather than interacting individual genes.[73] Studying modules interaction, thus, will provide us with a higher order view of biological system (see forest, not just trees) and understanding of causal relationship between functions.

In order to investigate the behavior of the pathways a dimension reduction procedure was proposed that transform expression values of all genes in a given module into one representative value for each sample. The approach implementing eigen vector was called eigengene.[74] Evaluation of statistical association between eigengenes tests a hypothesis of interaction between two pathways represented by corresponding eigengenes.

Differently from eigengene approach, multiple methods were proposed to calculate enrichment of links between members of separate pathways to identify crosstalking pathways based on diverse types of interactions such as protein interactions, co-expression etc..[75-77]

Once a relationship between modules/pathways has been established the next question is which nodes/genes are responsible for the interaction. Although multiple genes could act as mediators of interaction between two pathways, their relative importance can be different. Few approaches have been developed to find which nodes are critical for crosstalk between different modules in a network. Multiple sources of data are integrated to identify interactions between cancer-related pathways and key regulators are identified (genes that significantly altered for at least one molecular level) mediating those interactions.[75]

We have developed an approach that allows us to identify nodes in a network responsible for interactions between modules that potentially correspond to genes regulating crosstalk between pathways represented by these modules. Different from above discussed methods, this approach does not use any external information but is rather based upon topological properties of nodes in networks.

If we assume that the connections in network can be bi-directional and represent the same speed of information flow, then speed of information flow is controlled by the shortest path between two nodes. Considering the case of inter-pathway cross talk, the genes that are involved in shortest paths should be more important in controlling perturbation from one pathway to another. Several centrality measures have been proposed to evaluate the importance of nodes in a network.[78] Among those, betweenness centrality measures the importance of a node in acting as a bridge between any nodes within a network.[60] We modified standard betweenness centrality to adapt to the case of interaction between two defined subnetworks and to specifically address the question of which nodes belonging to "subnetwork 1" have a higher probability to be "bottlenecks" in the transfer of signal to the nodes in "subnetwork 2" and vice versa. For this metrics, the shortest paths are calculated only between nodes of two subnetworks and not between any nodes within a network.

We call this measure *bi-partite betweenness centrality*.

This metrics is calculated given by the following equation:

$$g(v) = \sum_{s \neq v \neq t} \frac{\sigma_{st}(v)}{\sigma_{st}}$$

Where $s$ belongs to "subnetwork1" and $t$ belongs to "subnetwork2", $\sigma_{st}$ is the total number of shortest paths from node $s$ to node $t$ and $\sigma_{st}(v)$ is the number of those paths that pass through vertex $v$ (node for which the metrics is calculated). Thus, this measurement represents the importance of a node in mediating information flow between two connected modules in a network. In our recent work we found that this approach allows finding not only "bottlenecks" of interaction between different pathways within the same organism but even microbial genes critical for mediating interaction between gut microbiota and their host (unpublished result).

*Revealing function of individual node in the network*

While most of our knowledge about gene functions is based on detailed and thorough gene-centered laboratory research, there are still genes whose functions have been less studied; network biology offers a novel way to infer functions for such genes. It uses an idea that genes that are located closely in a network may share a function. This principle is frequently called "guilt by association".[79] There are two major approaches that implement "guilt by association" for prediction of node function. The first is the so-called "direct approach". Although there are few slightly different methods using this approach (neighbor counting, graphic algorithm, probabilistic methods), they all assign a function to a node based on the functions of its direct neighbors.[80-82] The second approach, the "modular" approach, is to guide the assignment of a function to a gene by the collective function of other genes that belong to a given module in which the investigated gene is located.[48,83]

Besides identifying functions of individual nodes, generation of new ontology systems based on networks or pairwise similarities were proposed.[84] Interestingly, besides demonstrating a high level of consistency with existing ontologies, they provide solutions for situations when standard approaches(i.e. knowledge-based approaches) fail to reflect comprehensive biology.[85] Indeed, some terms/categories that were missing in the standard Gene Ontology (GO) and inferred by a network approach were submitted to the GO Consortium and incorporated into the ontology.[84,86]

*Networks cross-species conservation*
An important facet of network interrogation is the assessment of evidence for network function. Just as cross-species comparison is a core strategy for elucidating novel protein function (e.g., BLAST), cross-species comparison of network structure can reveal functions for network subgraphs that might not have been evident from sequence-level conservation of individual network components. In practice, subgraphs of the novel network (and in some approaches, constituent protein sequences) are used as keys to search for structural and component-sequence similarity to subgraphs in another species by searching for parsimonious subgraph-to-subgraph mappings (called a local network "alignment"). Alternatively, gene coexpression networks from two species can be compared in their entirety, to obtain a global alignment. A successful alignment enables all available functional annotations in the orthologous subgraph to bring to bear on the functional interpretation of the novel network's subgraph. Various local and global network alignment algorithms have been proposed, including NetworkBLAST,[87] PINALOG,[88] IsoRankN,[89] and the Narayanan-Karp[90] and Hodgkinson-Karp[91] algorithms.

**Different biological problems and some perspectives**

Some biological questions that can be addressed within the framework of network analysis remained beyond of the scope of this review. For example, one can try to evaluate number of nodes needed to be perturbed in order to achieve a transition from one state of biological system to another. This measure of network controllability[92] (number of needed nodes), although seemingly theoretical, can have very practical implications. On one hand, if a few nodes can govern a regulatory network modeling a disease, a gene perturbation/ gene silencing approach can be a good strategy for treatment.  On the other hand, if a large proportion of nodes in a network have to be modified in order to achieve recovery than a different pharmaceutical strategy using compounds that can simultaneously affect multiple molecular targets should be followed. Furthermore, some mathematical properties observed in biological networks such as "small world", "scale-free", "assortative mixing"[93] and several others[94] warrant further investigation to comprehend what types of environmental pressures led to selection of these properties during evolution and how they contribute to fitness and resilience of biological systems.

**In conclusion**
In this review we have described how network analysis can help to answer different questions commonly asked in biological research. We have also provided a detailed algorithm for this analysis including approaches employed by our group as well as frequently used in network-biology community.

## Acknowledgements
We thank Khiem Lam for editing.
AM is funded from NIH (AI109695; AI107485); AY supported by FAPESP (grant 2013/24516), LT supported by FAPESP (grants 2013/14722-8 and 2013/06223-1).

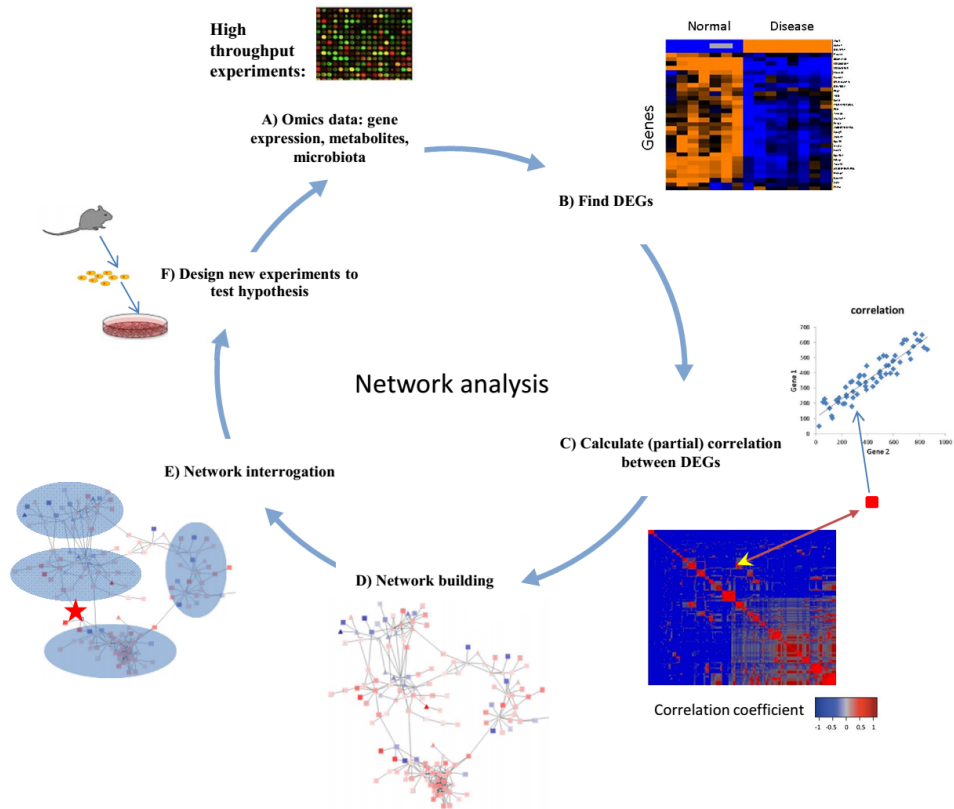

Figure 1. Workflow of network analysis. (A) Network analysis starts from data obtained from high-throughput experiments such as microarray experiments detecting expression of genes in samples. (B) Differentially expressed genes are found between two states of a system (e.g. normal vs. disease). (C) Correlations of DEGs based on their expression values are calculated to detect regulatory relationship among them. (D) Significant correlations suggest connections between DEGs and are used to generate a network of DEGs. (E) Network interrogation is performed to detect modules, key regulators, and functional pathways that are important for state transitions. (F) Based on the findings from network interrogation, new hypotheses are generated and can be tested in newly designed experiments. Data from new experiments could also be subject to further analysis.

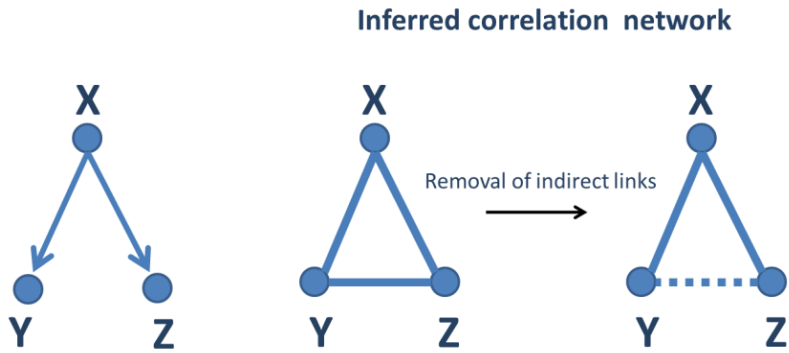

Figure 2. Removal of indirect links. As a demonstration, gene X can regulate the expression of both gene Y and Z. But there is no direct regulatory relationship between gene Y and Z. From the calculation of correlation of expression levels of three genes, correlations between gene X and Y, Z are observed as expected. However, gene Y and Z are also significantly correlated since they are both directly regulated by gene X. This correlation from common cause is called indirect link and can be removed by techniques like partial correlation, generating network reflecting regulatory relationships.

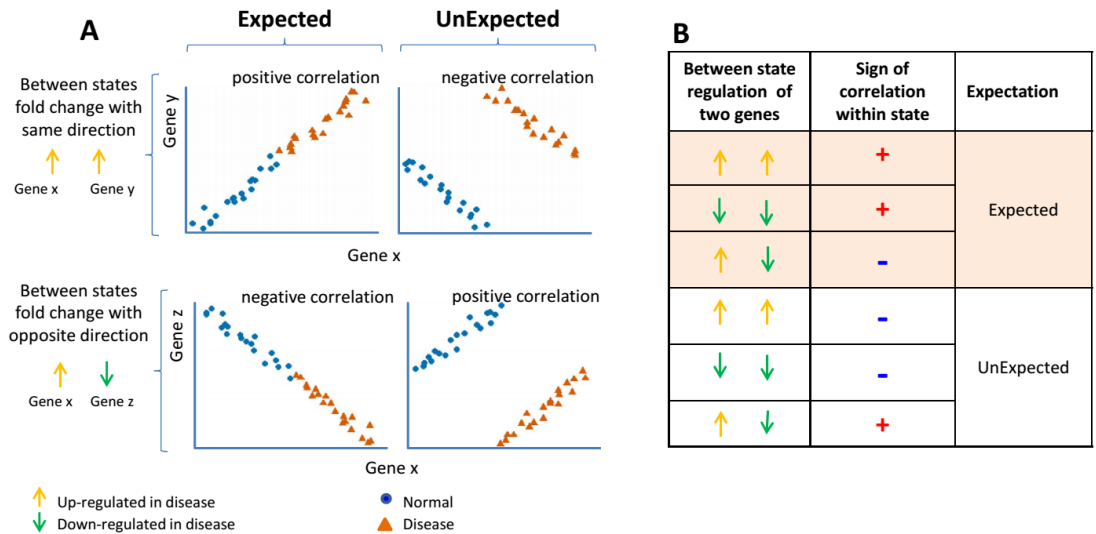

Figure 3. Illustration of expected and unexpected correlations. (A) When expression of two genes (gene x and gene y) are regulated toward the same direction when comparing two states, e.g. both upregulated in disease (upper two panels), we should expect their expression levels to be positively correlated within each state if there exists regulatory relationship between gene x and gene y. When two genes are oppositely regulated when transiting from normal to disease (in the lower two panels, gene x is upregulated while gene z is down regulated), we should expect negative correlation between those two genes in each state. (B) Different combinations of between states and sign of correlations used to define expected or unexpected correlation.

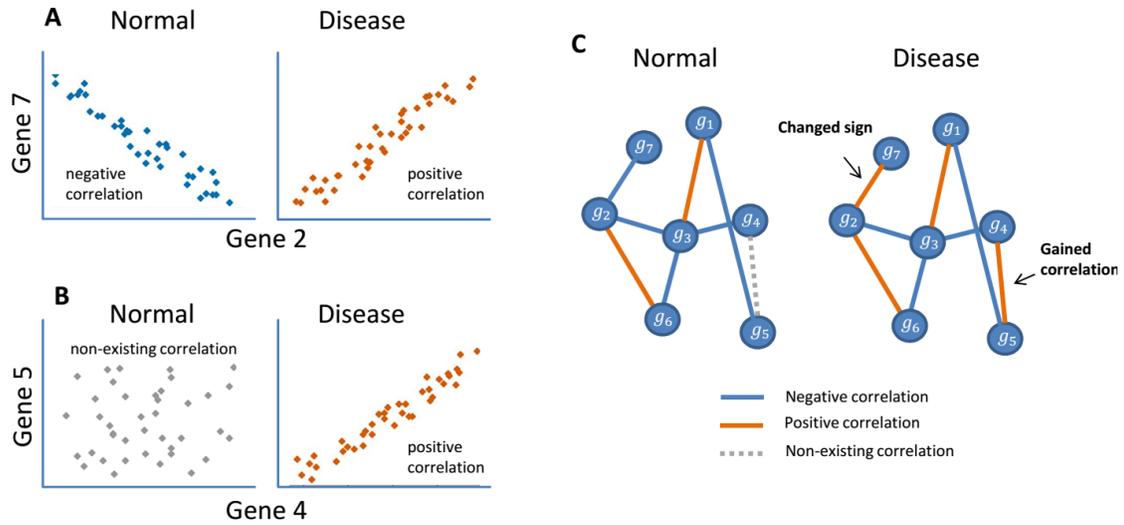

Figure 4: (A) Gene 2 and gene 7 correlate with each other in both normal and disease conditions, but the sign of correlation coefficient are opposite. (B) In normal condition there is no correlation between gene 4 and gene 5, but they gain positive correlation when the biological system transitioned to disease. (C) Example of visualization of the a network transitioning between normal and disease conditions. Red lines represent positive correlation, blue line represent positive correlations and dotted gray lines represent non-existing correlations in one condition that strongly appear in the other condition (on this case, becomes positively correlated).

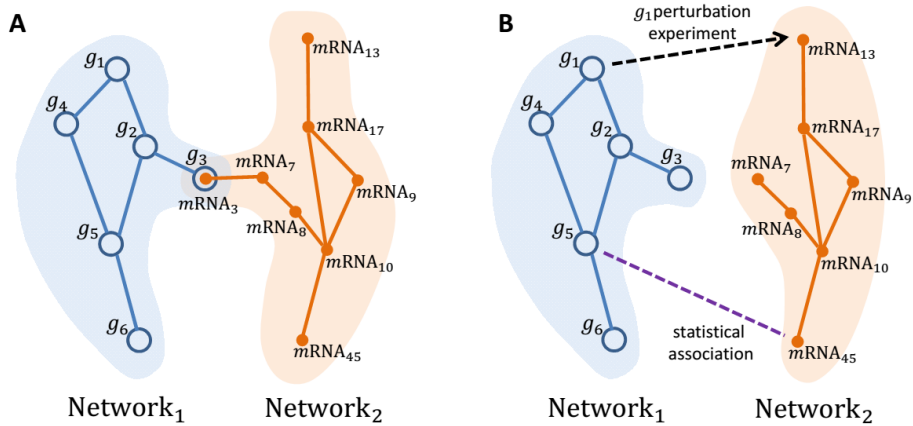

Figure 5. Data integration for inter-omics network. (A) Networks are constructed from different data types (e.g. network 1 for gene genetic interaction network and network 2 for mRNA coexpression network). These two networks then can be integrated into one network by overlapping the nodes that are correspondent between two networks (e.g. gene 3 and its transcript mRNA 3 are merged into one node).

(B) In another two types of integration, links are created between nodes by different evidence of "interaction", either experiment proved relationship (e.g. knockout of gene 1 altered the expression level of mRNA13) or statistical association between features of two nodes (e.g. gene 5 and mRNA45).

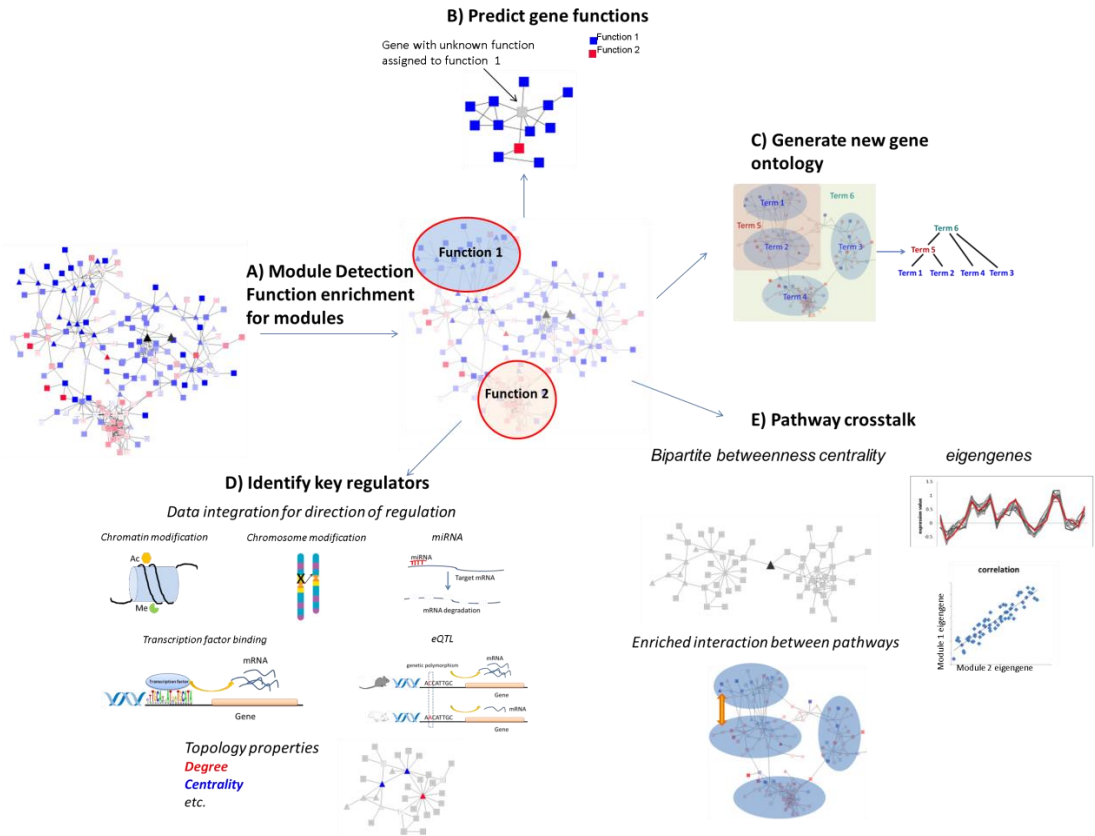

Figure 6. Network interrogation. (A) Densely connected sub-networks (modules) are detected and enriched functions of those modules are detected. (B)Genes with unknown function (gray) can be annotated based on the function of its neighbors in the network or the functions of the genes in the same module. (C) New gene ontologies can be generated by analyzing the hierarchical organization of gene clusters. (D) Multiple data types can be integrated to help infer direction of regulation and identify key regulators based on their network topological features. (E) Cross talks between pathways can be studied by extracting Eigengenes or analyze enriched interactions between networks. Key regulators for pathway cross talk can also be identified based on their between-module topology properties.

# Supplementary text

## Algorithm for calculation partial correlations.

Implementation of partial correlation is straightforward in R using the function `cor2pcor` from package "`corpcor`". The input of this function is a correlation matrix, which should be a "positive definite", a mathematically required property. However, in the omics data it is common that the correlation matrix is not positive definite because thousands of variables are measured in tens or hundreds of samples. Thus, to apply the inverse method we should make the estimation of the covariance matrix positive definite. For this, we use shrinkage estimation for covariance matrix which is implemented in R in the same package "`corpcor`".

```
>C = cor(X)
>C = make.positive.definite(C)
>C = cor2pcor(C)
```

## Algorithm for Meta-analysis scheme

1. Select only genes (or pairs of DEGs in case of networks) with the same direction of difference of mean (or correlation in case of networks) throughout all data sets. Each gene or gene-gene correlation should pass a certain p-value threshold. This threshold controls for heterogeneity between datasets.

2. For each possible gene (or pair of DEGs in case of networks), calculate the Fisher p-values using the following function created in R: let pvalue be the matrix where columns are studies and rows are genes, then

```
>pv.meta.analysis=function(pvalue){
>sum = log(pvalue[, 1]) # calculate fisher statistics
>for(j in 2 : ncol(pvalue)) sum = sum + log(pvalue[, j])
>t=-2*(sum)
>pv_fish=1-pchisq(t, 2*ncol(pvalue))  # calculate fisher pvalue
>return(pv_fish)}
```

Compute FDR over the obtained vector of Fisher p-values using the R function

```
>pv=pv.meta.analysis(pvalue)
>pv.fdr=p.adjust(pv, method = "fdr")
```

3. Select pairs with FDR less than a threshold.

## Algorithm for calculation p-values for DAPs

1. Calculate a matrix of pairwise correlation C1 and C2 for two groups of genes expression matrices X1 and X2 (number of columns is number of samples and rows are genes)

```
>C1 = cor(X1);   C2 = cor(X2);
```

2. We compute the p-values of difference of correlation using the function `pv.dif.cor.pearson` in R package "`psych`". Let n1 and n2 be the number of samples for corresponding groups (n1 is the number of columns in X1, n2 is the number of columns in X2).

```
>pv.dif.cor.pearson = function(C1,C2,n1,n2){
>if(!"psych" %in% installed.packages())
>install.packages("psych")
>library(psych)
># Convert correlations to z-scores
>z1 = fisherz(C1)
>z2 = fisherz(C2)
># Calculate vector of t-tests to compare
># correlations between classes
>fisher = (z1 - z2) / sqrt((1/(n1 - 3)) + (1/(n2 - 3)))
># Calculate raw p-values
>pv.dif.cor = 2*pt(-abs(fisher),Inf)
>return(pv.dif.cor)}
```

3. Compute FDR over the obtained vector of Fisher p-values using the R function. Select pairs with FDR less than a threshold.

```
>pv=pv. dif.cor.pearson(C1,C2,n1,n2)
>pv.fdr=p.adjust(pv, method = "fdr")
```

Table 1

| step | method -(statistics / mathematics) | software | link | ref |
|---|---|---|---|---|
| **network reconstruction** | | | | |
| normalization | quantile, lowess | BRB Array tools | http://linus.nci.nih.gov/BRB-ArrayTools.html | 95,96 |
| | quantile, lowess, etc. | package 'affy' in Bioconductor | http://www.bioconductor.org/packages/release/bioc/html/affy.html | 97 |
| | relevant mixture model framework | R package 'phyloseq' | http://joey711.github.io/phyloseq/ | 98 |
| finding DEGs | t-test | BRB Array tools | http://linus.nci.nih.gov/BRB-ArrayTools.html | |
| | different test statistics, choice with Bonferroni correction | IDEG6 | http://telethon.bio.unipd.it/bioinfo/IDEG6_form/ | 99 |
| regulation of genes | SVM | SIRENE | http://cbio.ensmp.fr/sirene/ | 100 |
| | semi-supervised learning; Logistic regression | SEREND | http://www.cs.cmu.edu/~jernst/Ecoli/ | 101 |
| | likelihood of mutual information | CLR | http://omictools.com/clr-s2342.html | 102 |
| | mutual information | ARACNE | http://wiki.c2b2.columbia.edu/workbench/index.php/ARACNe | 103 |
| | Mutual Information | MIDER | http://www.iim.csic.es/~gingproc/mider.html | 104 |
| | itemset mining | DISTILLER | request from authors | 105 |
| | Bayesian hierarchical clustering; conditional entropy | LeMoNe | http://bioinformatics.psb.ugent.be/software/details/LeMoNe | 106 |
| | Context Likelihood of Relatedness | inferelator | http://bonneaulab.bio.nyu.edu/networks.html | 107 |
| remove indirect links | partial correlation | corpcor | http://cran.r-project.org/web/packages/corpcor/index.html | 108 |
| | local partial correlation | | | 28 |
| | global silencing of indirect correlations | sILENCING | http://compbio.mit.edu/nd/ | 26 |
| | network deconvolution | network deconvolution | | 27 |
| weighted correlation network | pearson correlation | WGCNA | http://labs.genetics.ucla.edu/horvath/CoexpressionNetwork/Rpackages/WGCNA/ | 109 |
| differential co-expression | pearson correlation | CoXpress | http://coxpress.sourceforge.net/code.R | 110 |
| | pearson correlation | Dapfinder | http://exon.niaid.nih.gov/dapfinder/ | 111 |
| data integration | bicluster | cMonkey | http://bonneaulab.bio.nyu.edu/software.html#cmonkey | 112 |
| | itemset mining | DISTILLER | request from authors | 113 |
| meta-analysis | Fisher's combined probability test | metap' in software 'stata' | http://www.stata.com/support/faqs/statistics/meta-analysis/ | |
| | | OpenMeta | http://www.cebm.brown.edu/open_meta | |
| Visualization | | Cytoscape | http://www.cytoscape.org/ | 114 |
| | | Gephi | http://gephi.github.io/ | 115 |
| | | Circos | http://circos.ca/ | 116 |
| **network interrogation** | | | | |
| module finding | vertex weighting by local neighborhood density | MCODE | http://baderlab.org/Software/MCODE | 51 |
| | union of k-cliques | cfinder | http://www.cfinder.org/ | 52 |
| | Markov Cluster Algorithm | mcl | http://micans.org/mcl/ | 53 |
| function analysis/ gene set enrichment | Fisher's Exact | DAVID | http://david.abcc.ncifcrf.gov/summary.jsp | 55 |
| | Kolmogorov–Smirnov statistic modification | GSEA | http://www.broadinstitute.org/gsea/index.jsp | 117 |
| | Fisher's Exact | GoMiner | http://discover.nci.nih.gov/gominer/index.jsp | 118 |
| | Hypergeometric | GeneMerge | http://www.oeb.harvard.edu/faculty/hartl/old_site/lab/publications/GeneMerge.html | 119 |
| | Fisher's Exact | FuncAssociate | http://llama.mshri.on.ca/funcassociate/ | 120 |
| | dimension reduction (independent compoent analysis or fixed effect meta-estimate) followed by weighted pearson correaltion | ProfileChaser | http://profilechaser.stanford.edu/ | 121 |
| | hypergeometric test | Bingo | http://apps.cytoscape.org/apps/bingo | 56 |
| | Jaccard coefficient | EnrichmentMap | http://baderlab.org/Software/EnrichmentMap/ | 57 |
| | hypergeometric distribution | SubpathwayMiner | http://www.inside-r.org/packages/cran/SubpathwayMiner | 54 |
| identify Key regulators | network topology properties | cytoscape | tools::networkAnalyzer::Analyze network | 114 |
| | intramodular connectivity, causality testing | WGCNA | http://labs.genetics.ucla.edu/horvath/CoexpressionNetwork/Rpackages/WGCNA/ | 109 |
| pathway crosstalk | crosstalk enrichment | CrossTalkZ | http://sonnhammer.sbc.su.se/download/software/CrossTalkZ/ | 76 |
| | eigen vector | eigengene | http://labs.genetics.ucla.edu/horvath/htdocs/CoexpressionNetwork/EigengeneNetwork/ | 122 |
| gene function prediction | Bayesian network | MEFIT | http://mefit.joydownload.com/ | 123 |
| | fast heuristic algorithm from ridge regression | GeneMANIA | http://www.genemania.org/ | 124 |
| new gene ontology | hierarchical clustering | NeXO | http://www.nexontology.org/ | 125 |